# Valence band-satellite, temperature dependent magnetic and spectral study of α-Fe

Trishu Verma[1,*], Shivani Bhardwaj[1,], and Sudhir K Pandey[2,†]
[1]*School of Physical Sciences, Indian Institute of Technology Mandi, Kamand 175075, India and*
[2]*School of Mechanical and Material Engineering,*
*Indian Institute of Technology Mandi, Kamand 175075, India*
(Dated: December 8, 2025)

We investigate the influence of correlations and plasmonic excitation on valence band-satellite of α-Fe, along with magnetic and spectral properties as function of temperature. Coulomb interaction parameters are obtained by systematically employing various schemes in constrained random phase approximation (cRPA). This study identifies the presence of valence band satellite in Fe at ∼6 eV binding energy supported by (i) substantial incoherent spectral weight in the valence band spectra obtained from Density Functional Theory plus Dynamical Mean Field Theory (DFT+DMFT) and (ii) plasmonic excitations in the frequency range ∼6-8 eV suggested by $G_0W_0$ calculations. We note presence of significant contribution of temperature-dependent Pauli-spin susceptibility indicating competing degree of itinerancy. $e_g$ state shows a strong temperature driven non-Fermi-liquid behavior emerging near $T_c$. Our results reveal a high-temperature orbital-selective loss of coherence eventually leads to a orbital selective collapse of magnetization at $T_c$, suggesting a ferromagnetic phase characterized by strong correlation- and temperature- dependent spectral features.

## INTRODUCTION

Transition metals have partially filled, localized $d$-orbital. Therefore, correlation effects in these metals are so much vital that prediction of properties using theories based independent electron picture like Density Functional Theory (DFT), have been repeatedly proven ineffective[1–3]. The interplay of localization/correlation and itinerancy of electron in transition metals gives rise to variety of phenomena[4–6]. In particular, the $3d$ transition metals including V, Cr, Mn, Fe, Co and Ni, have been extensively studied for the electronic correlation effects due to broad spectrum of correlation strength and magnetic character displayed across the series. For example, bulk as well as surface superconductivity is observed for Vanadium [7, 8], with recently observed strain-induced magnetism[9]. Cr shows spin density wave in antiferromagnetic phase[10, 11]. While, Mn and Co are known for exhibiting multiple oxidation and spin states, leading to rich correlation-driven physics in their compounds leading to mixed valence, spin state transitions etc, which have been subject of continued interest[12–15]. β-Mn is found to exhibit frustrated magnetism induced heavy fermionic behavior[16].

Among these $3d$ transition metals Fe and Ni have long been used as prototypical systems for studying the electronic correlation effects in solids[17–19]. They are most abundant elements in earth's core[20], therefore, crucial to understand geomagnetism[21]. Their cubic crystal structure, room temperature ferromagnetism make them suitable materials for benchmarking theoretical methods for correlated electronic structure studies. Interestingly, Fe and Ni show remarkable similarities in several key features of correlation strength. For instance, both the metals show significantly reduced magnetic moments compared to their Hund's rule atomic moments i.e. from 4 (Fe) and 2 (Ni) $\mu_B$ to bulk values of 2.22 and 0.62 $\mu_B$, respectively[22]. This reduction reflects a similar coexistence of localized and itinerant $3d$ electrons in both systems. Furthermore, early dynamical mean field studies demonstrated ∼30% narrowing of the similar bandwidth of both systems[23–25]. Additionally, in literature a similar range of Hubbard $U$ value has employed for both the systems[18, 19, 23, 26]. However, despite similarities in various correlation strength signatures, their valence-band photoemission spectra shows striking qualitative differences that separate them. The famous 6 eV satellite in Ni is well recognized as a characteristic feature of its photoemission spectra and a fingerprint of strong correlations[27–29]. However, any such satellite feature is not reported in Fe[29–33], except a few resonant-XPS studies [34–37], which report a weak feature around 5.5 eV binding energy. Some researchers[33, 38] argued it to be caused by oxygen contamination by showing similar features could be enhanced when a clean sample is exposed to ample amount of oxygen. But Chandesris *et al* [35] claimed it to be genuine satellite feature. Since photoemission spectroscopy is a surface sensitive technique[39] and the traces of impurities (e.g. oxygen) could in turn mask low intensity satellite feature in background[40], hence experimental evidences may struggle to unambiguously comment on satellite in Fe. Therefore, despite the seemingly comparable correlation strengths, satellite peak of Fe is not experimentally well established leading to fundamental question of the origin and interplay of correlation effects and other losses indicated in the electronic spectra of these systems.

Moreover, recent insights into the satellite feature in Ni, where as opposed to the previous notion of solely correlation induced nature of the 6 eV satellite, suggest a comparable plasmonic loss contribution is inevitable in explaining its experimental intensity alongside the incoherent spectral weight[41]. This marks the presence of plasmonic excitations as a key factor for the satellite peak intensity in Ni. Since, the BCC Fe is also reported to exhibit plasmonic excitations in the 6-7 eV range[42], therefore, it becomes crucial to examine the relative contributions of the correlation effects and presence of plasmonic loss to establish whether the satellite feature is intrinsically missing in Fe or whether the experimental intensity is suppressed due to competing effects.

The origin of strong ferromagnetism in iron baffled researchers for many decades[43–46]. The advanced DFT+DMFT framework has proved to be quite successful in treating both the presence of itinerant quasiparticles and local multiplet physics in correlated systems on an equal footing[23]. Although, Fe remains ferromagnetic over a wide range of temperature with a high Curie temperature (∼1043 K), majority of the theoretical studies of Fe have been performed in the paramagnetic phase with largely density-density type of Coulomb interaction [45–49]. Thus a systematic analysis of the temperature-dependent evolution of spectral features in ferromagnetic phase is still missing. There have been numerous DFT+DMFT studies in the past done across wide range of $U$ values[18, 19, 23, 26, 49–53]. It is well know that the predictive power of DFT+DMFT largely depends on the Coulomb interaction parameters (Hubbard $U$ and Hunds $J$) used[54]. Therefore, a systematic study in this direction is needed to address the role of correlation effects in dictating its spectral weight distribution with realistic $U$ value to examine the magnetic and spectral properties in the ferromagnetic phase of Fe.

In this work, we employed different models of constrained RPA to determine Coulomb interaction parameters for Fe. A systematic investigation is carried out for the presence of valence-band satellite feature using a comparative analysis of correlation effects and plasmonic excitations. Further, the temperature dependence of the spectral features is given in the ferromagnetic phase till $T_c$. Our results reveal substantial degrees of itinerancy of the Fe 3$d$ electrons resulting from Pauli spin response, alongside an interesting orbital selective incoherence leading to the collapse of net magnetization in the system.

## COMPUTATIONAL DETAILS

DFT calculations are performed via full-potential (linearized) augmented plane-wave ((L)APW) + local orbitals (lo) method based code named WIEN2k [55]. The space group *Im-3m* with lattice parameter 2.829Å is used, optimized via fitting Vinet's [56] equation of states. The Generalized Gradient Approximation, given by Perdew-Burke-Ernzerhof (PBE) [57] is used as exchange correlation functional. The cRPA (to determine $U$ and $J$) and DFT+DMFT calculations are performed via GAP2 [58, 59] and eDMFT [60] code, re-spectively, both are interfaced with WIEN2k. In order to remove any parameter dependence, convergence test of properties is done with respect to important parameters like k-points, number of empty bands, etc. For cRPA calculation, k-mesh of $8 \times 8 \times 8$ is used, as, this $k$-mesh is found sufficient to converge Hubbard $U$ parameter within range of $\approx 0.1$eV. For DFT+DMFT, firstly, non-magnetic DFT is converged and then magnetism is treated with fully charge self-consistent DFT+DMFT. The convergence tolerance for self consistent field cycles is kept to be $10^{-4}$Ry/cell. Total number of Monte carlo steps in DFT+DMFT calculations are increased up to $10^9$, to reach the convergence of magnetic moment within 0.03 $\mu_B$. CTQMC (Continuous Time Quantum Monte Carlo)[61] impurity solver is used with Yukawa form[62] of exact double-counting scheme. For analytical continuation of complex functions in DFT+DMFT maximum-entropy method, implemented in eDMFT itself, is used. Single-shot $GW$ ($G_0W_0$) is performed using Elk code [63]. For analytical continuation of complex functions in $G_0W_0$, Padé approximation [64] is used. For $G_0W_0$ calculations, k-mesh of $12 \times 12 \times 12$ is used. We used full $W_0$ and diagonal $G_0W_0$ self energy matrix.

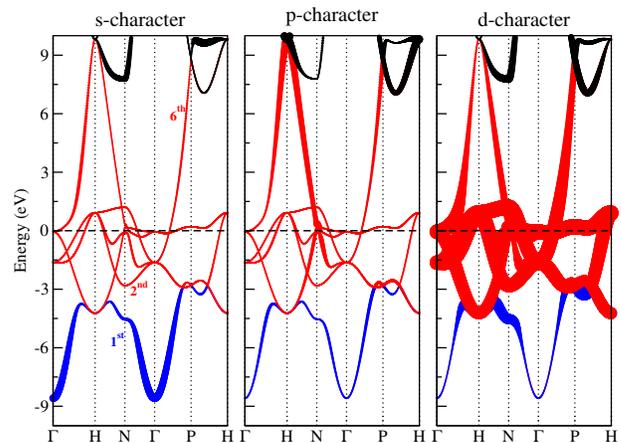

FIG. 1. Band-charater of Fe *s*-, *p*- and *d*- states in the non-magnetic phase. Blue, red and black colors show the $1^{st}$, $2^{nd}$ to $6^{th}$ and beyond $6^{th}$ bands respectively.

## RESULTS AND DISCUSSION

### Hubbard parameters and Magnetic properties

In this work, we have carried out the constrained random phase approximation (cRPA) calculations to estimate the Coulomb interaction parameters $U$ and $J$. The cRPA results are typically known to be highly sensitive to several parameters/factors in its formulation such as the choice of low-energy correlated subspace projectors, the number of bands constrained from participating in the screening process[65, 66]. In literature also, wide range of Coulomb parameters have been reported and used for $\alpha$-Fe, nicely summarized by Belozerov et al [49]. Considering all these aspects, we systematically varied bands constrained from the screening process and also employed the weighted approach to find the realistic $U$ and $J$ for Fe, obtained results are shown in table I, where $U_{full}$ and $U_{diag}$ represent the average of all and diagonal matrix elements of the effective Coulomb potential, respectively. For all cRPA calculations $1^{st}$ to $6^{th}$ bands (refer Fig. 1) were used to make maximally localized Wannier functions of $d$-character. To quantify the effect of screening from different bands we adopted different models. First we excluded screening from all the six bands which were used to construct Wannier functions. This resulted in large value of $U_{full}$ ($U_{diag}$) 8.35 eV (9.46 eV), refer (a) in table I. When, $1^{st}$ band was allowed to contribute in screening process as given in (b), the $U_{full}$ ($U_{diag}$) drastically reduced to half of the value obtained in case (a) i.e. 4.58 eV (5.63 eV). This shows the importance of screening from $1^{st}$ band which has majority of $s$- and $d$-character.

TABLE I. Coulomb interaction parameters(on-site $U$ and exchange integral $J$) for different models calculated via cRPA. $U_{full}/U_{diag}$ represents the average over all/diagonal elements of on-site interaction matrix.

|     | Constrained Bands | $U_{full}$ | $U_{diag}$ | $J$ |
| --- | --- | --- | --- | --- |
| (a) | 1-6 | 8.35 | 9.46 | 0.69 |
| (b) | 2-6 | 4.58 | 5.63 | 0.66 |
| (c) | 1-6,Weighted | 2.53 | 3.55 | 0.64 |
| (d) | Fully Screened | 0.30 | 0.84 | 0.34 |

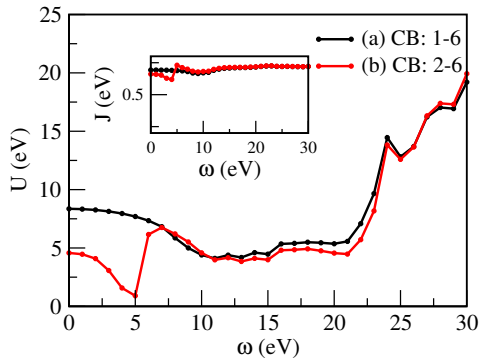

FIG. 2. $U(\omega)$ when constrained bands (CB) were $1^{st}$ to $6^{th}$ (black) and $2^{nd}$ to $6^{th}$ (red) i.e. model (a) and (b). Inset: $J(\omega)$

Next, frequency dependence of $U_{full}$ ($U(\omega)$) and $J$ ($J(\omega)$) is given in Fig. 2, corresponding to (a) and (b) models. In low frequency region we observe significant difference in $U(\omega)$ curves for (a) and (b) models, especially around 5 eV where model (a) assumes significant dip while (b) remains relatively constant. This could be regarded as signature of onset of plasmonic excitations as suggested by Sihi et al[53]. It should be noted that above 7 eV both models are in well agreement with each other, suggesting the screening from $1^{st}$ band is crucial at low frequencies. The absence of dip in model (a) where $1^{st}$ band is included in the correlated subspace and presence of dip in model (b) where $1^{st}$ band in excluded from the correlated subspace, suggests the contribution to the screening effects from the $4s$ electrons in leading to the plasmonic excitations in Fe.

Now, since $1^{st}$ and $6^{th}$ bands show significant mixture of $s$- and $p$-characters, respectively, entangled with substantial $3d$ character, a simple unweighted masking of transitions (as employed in cases (a) and (b)) would lead to inaccurate separation of the correlated $d$-subspace from the uncorrelated screening channels. Therefore in part (c) of Table I, we employed the weighted mask approach[19], since it is expected to provide more realistic treatment of screening by assigning weights to the transitions according to $d$ orbital character. The $U_{diag}$ by this approach was found to further reduce to 3.55 eV, in close agreement with values, earlier reported by Şaşıoğlu et al[19]. The calculated $U_{full}$, using the weighted approach comes out to be further reduced with respect to the value obtained in models (a) & (b). The significant reduction in $U_{full}$ and $U_{diag}$ values using the weighted approach indicates the importance of screening effects from $s$ and $p$-character electrons in the bands which were used to make the Wannier functions. In all these above models we find that $J$ value shows negligible change ($\simeq 0.66$ eV) and lies close to its atomic value[49, 67]. In contrast to the expectation that set of Coulomb interaction parameters ($U_{full}$ and $J$) obtained from weighted approach should be better representative parameters to perform realistic DFT+DMFT calculations, we find large underestimation of the low temperature magnetization in Fe with thus obtained on-site Coulomb interaction parameters i.e. $U_{full} = 2.53$ eV (Yukawa[62] calculated $J$ is suggested to be better choice for the DFT+DMFT calculations [41, 53, 68], therefore correspondingly calculated $J$ is used against $U_{full}$ value). Additionally in (d), we report the fully screened $U$ and $J$ parameters, where $U_{full}$ & $U_{diag}$ decrease to less than 1 eV i.e. 0.30 eV & 0.84 eV, while $J$ value reduces by almost half of the previous values.

As, mentioned Coulomb interaction parameters obtained via cRPA weighted approach underestimates low temperature magnetization. Therefore, we benchmarked $U$ ($J$ correspondingly obtained via Yukawa screening potential approach[62]) for low temperature magnetization. We note that $U = 4.09$ eV and $J = 0.90$ eV provide better agreement of low temperature magnetization value with the experimental results, therefore, for all further DFT+DMFT calculations we have used this set of $U$ and $J$. This set of $U$ and $J$, overestimates $T_c$ up to around 1800 K, around 1.7 times more than experimental value. This is consistent with previous single-site DFT+DMFT studies where calculated $T_c$ value was overestimated till 2500 K and between 1500 - 1800 K, using density-density and rotationally invariant, form of Coulomb interaction parameters, respectively[53, 69, 70]. Interestingly, the experimental $T_c$ value was reproduced well with $U = 2.43$ and $J = 0.69$ eV, but with significant underestimation of the saturation magnetization, by almost 21% of the experimental value (2.18 $\mu_B$)[22]. Subsequently, to validate whether the reduced temperature scaled magnetization curve can be obtained from the DFT calculations as proposed in Ref.[41], we present the results for the finite temperature magnetization curve ($M(T)$) obtained with DFT calculations along with the DFT+DMFT and experimental[22] curves (given in Fig. 3). DFT overestimated $T_c$ many times (7200 K), which was expected considering its mean field nature. However, when the temperature axis is scaled with the scaling factor obtained in such a way that $T_c$ predicted by DFT (i.e. 7200 K) coincides with the experimental value (scaling factor = 1043/7200= 0.14) a scheme proposed in Ref. [41], we find astonishingly good agreement with the experimental $M(T)$ curve till T~700 K. Above 700 K, a consistent slight underestimation is noted in the values till $T_c$. The DFT+DMFT curve provides good match of the magnetization values at temperatures till ∼600 K but deviates thereafter, causing large overestimation of critical temperature. The overestimation is largely attributed to the missing inter-site exchange interactions in single-site DMFT approach. Similar to DFT, we also provided scaled temperature axis for DFT+DMFT curve, with correspondingly deduced scaling factor (= 1043/1800 = 0.58). Interestingly, the temperature scaled DFT+DMFT curve also provides nearly good agreement with the experimental magnetization values in the temperature region below 500 K. Surprisingly, better description of experimental $M(T)$ curve with finite-temperature magnetization estimates via DFT by employing an appropriate temperature scaling factor in Fe suggests a competing itinerant nature of magnetism, as also proposed for Ni[41].

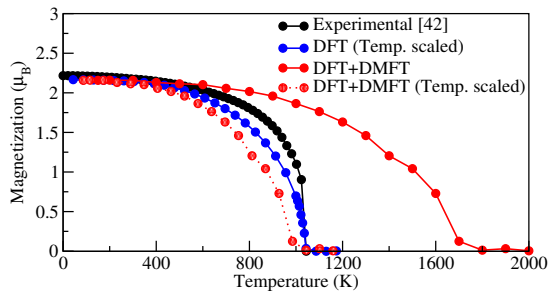

FIG. 3. Magnetization vs. temperature ($M(T)$) curve obtained via DFT and DFT+DMFT along with the experimental[22] $M(T)$ curve.

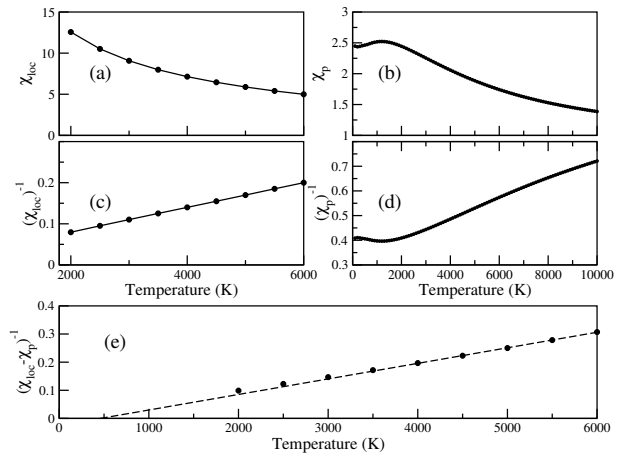

FIG. 4. Spin-susceptibility vs. temperature, (a) shows total local spin magnetic susceptibility, (b) shows the Pauli susceptibility. (c) and (d) shows inverse of local spin and Pauli susceptibilities, respectively (e) shows $\chi_p$ subtracted local magnetic spin susceptibility along with CW fit for temperature above 4000 K, extrapolated to calculate $T_c$. Units of susceptibility and inverse susceptibility are $\mu_B^2/eV$ and $eV/\mu_B^2$ respectively.

Further, we calculated temperature dependent local spin magnetic susceptibility ($\chi_{loc}$), shown in Fig. 4(a). We calculated Pauli susceptibility ($\chi_p$) directly via solving Kohn-Sham DFT Hamiltonian without considering exchange-correlation potential and computing the change in total magnetic moment with respect to external magnetic field (in the linear response range). The Pauli susceptibility thus calculated is shown in Fig. 4(b). Inverse of local ($\chi_{loc}^{-1}$) and Pauli susceptibilities ($\chi_p^{-1}$) are shown in Fig. 4(c) and 4(d), respectively. $\chi_{loc}^{-1}$ although have linear dependence above temperature 2000 K, but CW couldn't fit directly because of the possible presence of non-negligible $\chi_p$ contribution as found from the subsequent calculations. In contrast to the conventional notion of temperature-independent behavior of $\chi_p$, the curve here shows an observable temperature dependence, where the value reduces to 55% of its maximum value in the given temperature range. The $\chi_p^{-1}$ remains almost constant up to around 2000 K, followed by a Curie-Weiss like behavior towards high temperature region, as depicted in Fig. 4(d).

To understand the origin of Curie-Weiss like behavior in the $\chi_p$, we calculated the Fermi temperature of Fe using the density of states at Fermi-level following the textbook procedure[71]. Typically, such a Curie-Weiss like behavior can arise when a degenerate Fermi-gas approaches the Boltzmann regime at temperatures comparable to or larger than the Fermi temperature[72]. We find that the Fermi temperature for Fe comes out to be of the order of $10^7$ K, which is huge compared to the temperature scales used in this study. This suggests that electrons continue to obey the Fermi Dirac statistics. Hence, the emergence of the Curie-Weiss behavior can be attributed to the pre-localization effects[70]. The pre-localization effects can also be regarded to originate from the presence of flat band near the Fermi-level along the $N - \Gamma - P$ high-symmetric $k$-directions (refer to Fig. 1). At high temperature, value of $\chi_{loc}$ is very low therefore finite $\chi_p$ contribution becomes more significant. Considering also recently noted contribution of $\chi_p$ in Ni and here in Fe, it can be concluded that although both Ni and Fe have significant itinerant moments however, the local moment picture in Fe must be more robust than in Ni[70]. Next, the CW fit for $\chi_{loc}$ after subtracting the $\chi_p$ contribution is given in Fig. 4(e) and fitted CW law, for determining $T_c$, in the range of 4000 K to 6000 K. The $T_c$ calculated via this method are 460 K. We also note a significant dependence on temperature range on which CW law is fitted. For e.g. when CW law is fitted in temperature range 2000 K to 6000 K, obtained $T_c$ is 162 K.

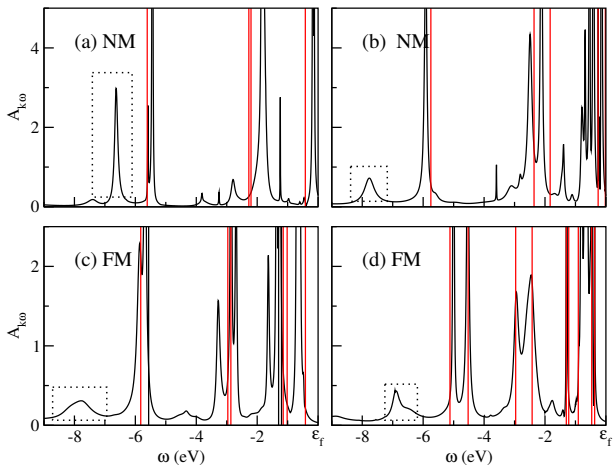

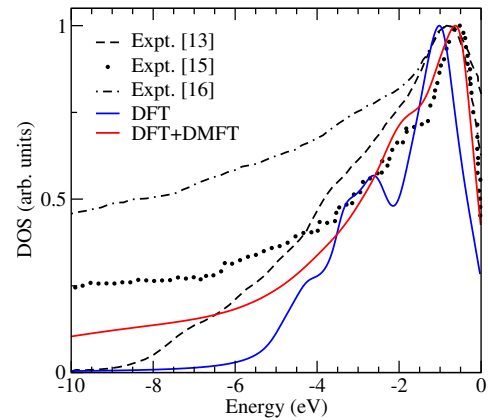

FIG. 5. Spectral function calculated via $G_0W_0$. (a), (b) are spectral function from NM calculation at $k = (0.167, 0.167, 0)$ and $(0.428, 0.036, 0)$ and (c), (d) are for FM calculation at $k = (0.167, 0.083, 0.083)$ and $(0.667, 0.333, 0.083)$, respectively. Straight red lines, parallel to ordinate-axis represents DFT eigenvalues.

FIG. 6. Calculated 3$d$-character DOS at 300 K (using DFT and DFT+DMFT, with lorentzian broadening of 0.55 eV) along with experimental XPS data obtained at photon energy of 1486.6 eV [31] and 1253.6 eV [29, 32].

### Investigation of valence band satellite and quasiparticle properties

Further, in order to confirm the presence of plasmon in Fe, we did $GW$ calculations for both nonmagnetic (NM) and ferromagnetic (FM) phases of Fe at 300 K. To rule out the features generated from the long range many body effects arising as a result of self consistency condition of $GW$ calculations we restricted our $GW$ calculations to single iteration i.e. single-shot $GW$ ($G_0W_0$). Since, the spectral weight due to plasmonic loss in $k$-integrated DOS can not be unambiguously identified, therefore we studied spectral function ($A_k(\omega)$) for individual $k$-points. There were only few $k$-points for which $A_k(\omega)$ had finite incoherent weight in the range of 6-8 eV. $A_k(\omega)$ for two such $k$-points from each phase are presented in Fig. 5. This further validate the plasmonic loss present in 6-8 eV from the Fermi level ($\varepsilon_f$). Moreover, we also calculated the plasma frequency ($\omega_p$) for both phases. For, NM and FM phases, $\omega_p$ came out to be 6.15 eV and 7.27 eV, respectively, which is in good agreement with recently reported value[42]. This difference was expected because in FM phase RPA could not correctly account for fluctuating moment at finite temperatures. Similar difference between $\omega_p$ of the two phases was reported earlier for Ni [41]. Consistency of $\omega_p$ values within the dip $\omega$ dependent Coulombic interaction parameters and the finite spectral weight corresponding to the ∼6 eV plasmonic loss confirms the presence of plasmonic excitation in BCC Fe.

Here the results indicate the presence of 6 eV plasmonic loss feature in the valence band spectrum of Fe. Notably, similar presence of substantial spectral weight of plasmonic excitation has recently been reported in case of Ni, where plasmonic loss intensity together with the correlation-induced incoherent weight contributions account for the experimentally observed 6 eV satellite in Ni. Therefore examining the spectral weight distribution in the photoemission spectra (PES) of Fe will be crucial for further insights into the differences of electronic correlations between Fe and Ni and for clarifying the microscopic origin of the emergence or suppression of a satellite features in these correlated transition metals. We obtained the density of states (DOS) at 300 K from the DFT+DMFT calculations, which is given in Fig. 6 along with the PES and DFT DOS. The experimental PES plotted here are obtained with Al $K_\alpha$ (1486.6 eV) [31] and Mg $K_\alpha$ (1253.6 eV) [29, 32] X-ray source. Since the cross-section of the $3d$ states is more than three times of the $4s$ states [73] for both photon sources, the XPS spectrum is compared with the $3d$ DOS obtained from the DFT and DFT+DMFT calculations. Also, to take into account the instrumental broadening [31, 73], the calculated DOS was convoluted with the Lorentzian of 0.55 eV after multiplying with the Fermi-Dirac distribution function. It could be observed from the PES that there is a peak near binding energy 1 eV which is predicted with both DFT and DFT+DMFT but DFT+DMFT gives its position closer to PES. The DFT+DMFT not only generates feature's line-shape but also the yields position of the plateau in good agreement with experiment, which is not produced by DFT. Beyond 5 eV of binding energy DFT DOS starts to drops rapidly. After 6 eV it gives negligible DOS. However, there is finite incoherent weight in DFT+DMFT DOS ranging from 6-8 eV and beyond.

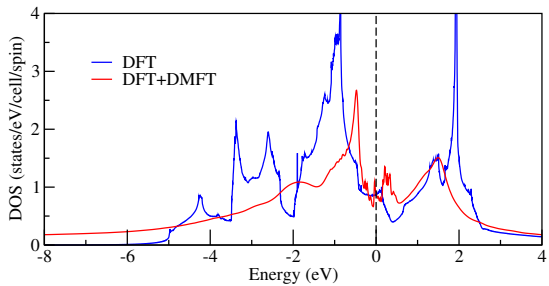

FIG. 7. Density of states for $d$-character calculated by DFT and DFT+DMFT at 300 K, without any convolution/broadening.

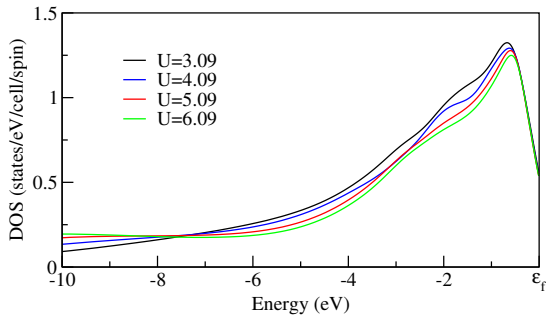

FIG. 8. $d$-character DOS for different $U$ and fixed J(=0.9 eV), calculated by DFT+DMFT at 300 K.

To assess, that the incoherent weight obtained in the energy region beyond 6 eV in DFT+DMFT curve was not due to convolution or broadening effects we studied the raw DOS obtained, along with the DFT raw DOS, as shown in Fig. 7. It could be observed that even though DFT's spectrum completely dies out above 6 eV but DFT+DMFT shows considerable spectral weight. Which indicates incoherent weight observed in the region is genuine feature, resulting from the electronic correlations. Further to verify whether the incoherent weight in the region beyond 6 eV arises from the choice of the $U$ parameter, we systematically varied $U$ and re-examined the spectra. The incoherent weight component around plasmonic loss frequency region (6-8 eV) remains constant across the studied range of $U$ values (3-6 eV), indicating that its a genuine robust feature of the electronic spectrum and not an artifact of the interaction parameter strength as illustrated in Fig. 8.

We note clear presence of finite spectral weight in the 6-8 eV binding energy range which coincides with the expected plasmonic loss energy in Fe. This persistence of the incoherent spectral weight in its DOS irrespective of the choice of $U$ in principle strongly suggests the plausibility to host a satellite feature $\sim$6 eV binding energy as in the case of Ni. However, an important distinction can be noted that the plasmonic loss intensity in the $k$-integrated spectral function density would be suppressed considerably since the presence of plasmonic excitations are found for only few $k$-points -thus resulting in weak intensity $\sim$ 6 eV region in comparison to Ni. Nevertheless, given the finite incoherent weight and plasmonic excitations in the energy region $\sim$ 6 eV the presence of low intensity satellite feature in this energy range cannot be ruled out in Fe. Although, in contrast to Ni, the existence of valence band satellite in Fe has been quite extensively debated in several previous experimental studies[34, 35, 38, 74, 75]. Moreover, there have been divided reports on the experimental evidence of such a feature in Fe, for instance, early resonant XPS measurements by Chandesris *et al*[34] showed a satellite in Fe around 5 eV, whereas several later studies attributed the observed valence band satellite like feature to the oxygen contamination. However, subsequent work by Chandesris *et al*[35] again demonstrated the presence of satellite even for the clean surfaces, suggesting contamination alone cannot account for the observed feature. Considering these studies its necessary to note certain aspects of obtaining the EES experimentally- firstly, the photon energies used for obtaining the EES are relatively low thus limiting them to only the top few atomic layers, making it to detect even minimal oxygen adsorption which can partially mask the genuine intrinsic satellite signal. Secondly, in PES of Fe, the position of the peak in the unoccupied region lies $\sim$2 eV above the Fermi-level and therefore to exhibit strong resonance, Fe would require a specifically large photon energy ($\sim$54 eV corresponding to the $3p \to 3d$ transition) to enhance the satellite feature in the resonant XPS. In fact, the experimentally reported resonant XPS[33, 34] with photon excitation energies in the range 54-54.5 eV show a visible enhancement around 5.5 eV binding energy consistent with the presence of the satellite feature seemingly superimposed on residual oxygen induced background. Moreover, in the experimental EES of Fe, a non-negligible spectral weight can be consistently observed to exist at 5-8 eV binding energy in the available experimental studies[29–33]. This high binding-energy residue aligns with the plasmonic loss energy and the correlation induced incoherent weight in the theory, thereby suggesting a plausibly suppressed satellite intensity which remains less pronounced than that observed in Ni due to remarkably less intense plasmonic excitations.

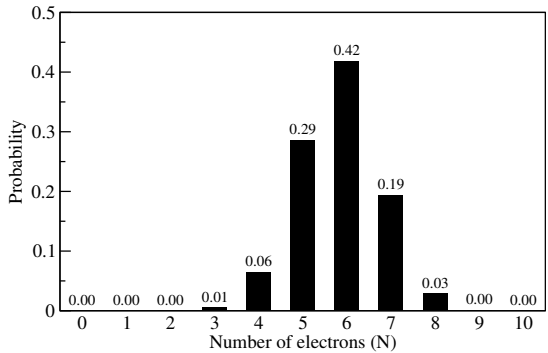

FIG. 9. Probability distribution of electronic configuration for $3d$ orbital at T = 300 K.

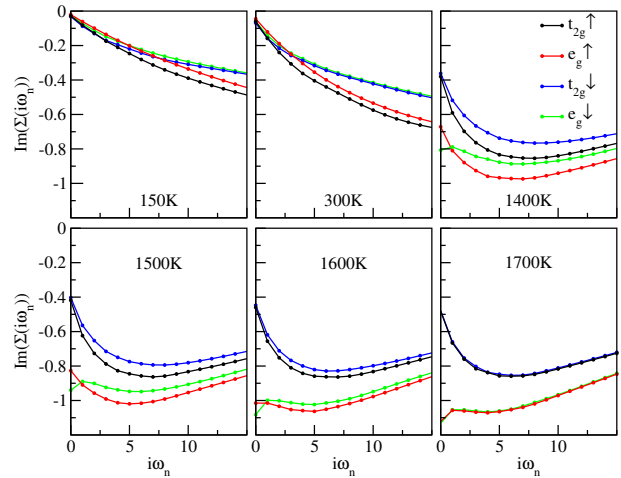

FIG. 10. Orbital and spin resolved imaginary part of self energy ($\text{Im}(\Sigma(\iota\omega_n))$) as function of Matsubara frequency, at different temperatures below $T_c$.

TABLE II. Orbital and spin resolved occupancy at some temperatures. Note: $\uparrow$ and $\downarrow$ represents up and down spin states respectively.

| Temperature (in K) | $t_{2g}\uparrow$ | $e_g\uparrow$ | $t_{2g}\downarrow$ | $e_g\downarrow$ |
|---|---|---|---|---|
| 150 | 2.35 | 1.64 | 1.24 | 0.59 |
| 1400 | 2.13 | 1.41 | 1.48 | 0.80 |
| 1700 | 1.83 | 1.12 | 1.78 | 1.08 |
| 1800 | 1.81 | 1.10 | 1.80 | 1.10 |

To discuss the intensity of satellite we also calculated configuration probabilities for $3d$ orbital at 300 K via DFT+DMFT, shown in Fig. 9. The probability distribution was found to be temperature independent, over the studied range. In the calculated histogram several initial state configurations i.e. $3d^5$, $3d^6$ and $3d^7$ -show finite and comparable probabilities. Where $3d^6$ is the most probable configuration. The $3d^7$ is approximately half of the most probable configuration and $3d^5$ also carries 29% weight. As a result not only the primary $3d^6 \to 3d^5$ channel but also the closely lying multiplet channels are expected to contribute to the photoemission spectrum with appreciable intensity. Since these multiplets lie close in binding energy their individual contributions overlap strongly which leads to broad plateau like satellite rather than sharp, well separated peak, consistent with the extended incoherent weight over a large energy region 6 eV and beyond in the DFT+DMFT spectra.

The aforementioned discussion shows that the suppression of the satellite-like feature in the EES of Fe arises not from the fundamental absence of correlation-induced spectral weight, but from the comparatively weak plasmonic loss contribution, thus diminishing the overall observable intensity of the satellite feature in both experiment and theory.

Further we studied the imaginary part of self energy, $\text{Im}(\Sigma(\iota\omega_n))$, in the FM phase of Fe at different temperatures (given in Fig. 10). To avoid any analytic continuation caused noise, self energy sampled in Matsubara frequencies ($\iota\omega_n$) are directly used[76]. At low temperatures (in the range 150-300 K), both $t_{2g}$ and $e_g$ orbitals behave like Fermi-liquid showing linearly vanishing curve towards zero $\iota\omega_n$. At high temperatures, around 1400 K and above, $e_g\downarrow$ shows a persistent kink at zero $\iota\omega_n$. The $e_g$ orbital is said to deviate form Fermi-liquid behavior based on this kink in some earlier studies[45, 49, 70] but we could not find any study in which temperature evolution of spin resolved self energy is discussed. At 1500 K emergence of orbital selectivity could be observed. Such a kink signals loss of quasiparticle coherence. This downturn/kink also starts to appear in the $e_g\uparrow$ orbital above 1600 K. This behavior shows that the system gradually becomes orbital selective, where different orbitals lose coherence at different temperature. This interpretation is strongly supported by the temperature-dependent orbital occupations given in Table II. The temperature dependent occupations for $t_{2g}$ and $e_g$ orbitals indicate the collapse of their individual exchange splitting, leading to net collapse of total magnetization at 1800 K. This is consistent with a Hund's induced collapse of orbital-resolved exchange splitting. Notably, the collapse of Hund's like exchange-splitting for both the $t_{2g}$ and $e_g$ orbitals occur simultaneously at $T_c$ despite different coherence scales. Interestingly, such an orbital-dependent loss of coherence has also been reported earlier in Fe-based correlated materials[77], where Hund's physics is known to induce orbital selectivity at finite temperature[78, 79]. Therefore, the kink observed here can be ascribed as a precursor to orbital selectivity building near $T_c$.

## Conclusion

In conclusion, we revisited $\alpha$-Fe to study the temperature dependent evolution of its magnetic and spectral properties. Our study reveals the presence of valence band satellite in Fe around 6 eV region which is established from the presence of significant incoherent weight and presence of plasmonic excitations around 6-8 eV confirmed from the constrained RPA and $G_0W_0$ calculations. We further explain the plausible reasons for its suppressed experimental visibility, by attributing it to the weak intensity of plasmonic excitations and overlapping multiplet contributions ($3d^5$, $3d^6$ and $3d^7$) smear the intense feature into broad plateau of incoherent spectral weight. This explains why only few of the several experimental studies have reported the valence band satellite $\sim$5.5 eV binding energy despite its fundamentally intrinsic origin.

We showed that Hubbard $U$ estimated from the weighted approach of cRPA fails to account for the low temperature magnetization behavior of Fe. We also calculated temperature dependent magnetization and spectral properties. Interestingly, with temperature axis scaled, DFT gives good match with temperature dependent magnetization curve, suggesting competing itinerant nature of magnetism. We note significant Pauli spin response to the magnetic spin susceptibility in the paramagnetic phase above $T_c$ resulting from competing itinerant nature of magnetism along with a nice Curie-Weiss fit of the inverse local spin susceptibility characterizing the local moment formation. Furthermore, the temperature dependent self-energy variation shows an orbital-dependent loss of coherence towards approaching $T_c$ and a Hunds driven collapse of exchange splitting leading to collapse of net magnetization at $T_c$.

## Acknowledgments

We acknowledge National Supercomputing Mission (NSM) for providing computing resources of 'PARAM Himalaya' at IIT Mandi, which is implemented by C-DAC and supported by the Ministry of Electronics and Information Technology (MeitY) and Department of Science and Technology (DST), Government of India. The research of "Trishu Verma" is funded by University Grants Commission (UGC) with Fellowship No. 231610046039.

---

* trishumahender@gmail.com
† sudhir@iitmandi.ac.in